# Merging Control Strategies of Connected and Autonomous Vehicles at Freeway On-Ramps: A Comprehensive Review

Jie Zhu [1], Said Easa [2], Kun Gao [*,1]

[1] Department of Architecture and Civil Engineering, Chalmers University of Technology, 41296 Gothenburg, Sweden

[2] Department of Civil Engineering, Ryerson University, Toronto, Ontario, Canada

**Abstract**

On-ramp merging areas are typical bottlenecks in the freeway network, since merging on-ramp vehicles may cause intensive disturbances on the mainline traffic flow and lead to various negative impacts on traffic efficiency and safety. The connected and autonomous vehicles (CAVs), with their capabilities of real-time communication and precise motion control, hold a great potential to facilitate ramp merging operation through enhanced coordination strategies. This paper presents a comprehensive review of the existing ramp merging strategies leveraging CAVs, focusing on the latest trends and developments in the research field. The review comprehensively covers 44 papers recently published in leading transportation journals. Based on the application context, control strategies are categorized into three categories: merging into sing-lane freeways with total CAVs, merging into sing-lane freeways with mixed traffic flows, and merging into multilane freeways. Relevant literature is reviewed regarding the required technologies, control decision level, applied methods, and impacts on traffic performance. More importantly, we identify the existing research gaps and provide insightful discussions on the potential and promising directions for future research based on the review, which facilitates further advancement in this research topic.

**1 Introduction**

On-ramp merging is critical for freeway traffic operation since the cut-in maneuvers of ramp vehicles impose frequent disturbances on the mainline traffic flow and cause various problems. These include traffic oscillations, increased fuel usage and emissions, safety concerns, and recurrent traffic congestions (Cassidy et al., 1999, Mergia et al., 2013, Srivastava et al., 2013, Han et al., 2018, Wang et al., 2019, Larsson et al., 2021, Ali et al., 2021, Zhang and Yang 2021). However, with vehicle communication and autonomous driving technologies, new possibilities exist to prevent or mitigate such adverse traffic effects in the ramp merging areas. Traditional traffic management approaches such as ramp metering (Papageorgiou et al., 1991, Papageorgiou et al., 1997, Ahn et al., 2007, Papamichail et al., 2010), variable speed limits/mainline metering (Carlson et al., 2010, Jin et al., 2017, Zhang et al., 2017, Peng et al., 2021, Lu and Liu 2021, Chen et al. 2021), and hard shoulder running (Haj-Salem et al., 2014, Li et al., 2014). Unfortunately, these approaches can only control the traffic at an aggregated level. The emerging connected and autonomous vehicles (CAV) provide an opportunity to regulate the behaviors of individual vehicles and facilitate advanced cooperation and coordination in the on-ramp merging areas.

Many strategies based on the communication and automation capabilities of CAVs have been developed over the last decades, devoted to facilitating the merging/ lane-changing maneuvers at freeway on-ramps. Though sharing the common objective of improved ramp merging operation, the existing strategies present substantial differences in many aspects, for example, the required vehicle technologies (connected, autonomous, or connected and autonomous), the required level of automation (fully automated, partially automated, or driver-assisted), the direction of control (longitudinal, lateral, or both), the method of control (optimal control, feedback control, or others), and the type of control (centralized or decentralized). The prior efforts have been reviewed by Scarinci et al. (2014), Bevly et al. (2016), and Rios-Torres et al. (2017b). As the initial attempts to coordinate CAVs at on-ramp merging, the prior strategies mainly focus on interactions between a few individual vehicles. Their benefits are only demonstrated in simple use cases. Further, assumptions on traffic compositions and freeway layouts are usually simplified in these studies. In recent years, significant advances have been made in CAV ramp merging. These include (1) developing strategies targeting improvements at the continuous traffic flow level, (2) developing strategies for the traffic conditions where CAVs and Human-Driven Vehicles (HDV) coexist, and (3) developing strategies for multilane freeways where the free lane changes between mainline lanes may affect the merging traffic. Despite the significant progress made, an up-to-date review covering these latest developments is missing to the authors' best knowledge.

This paper conducts a thorough review of the cooperation/coordination strategies that facilitate freeway on-ramp merging using CAVs, focusing on the latest developments in this field. Based on the review, we identify the existing research gaps in CAV ramp merging and discuss the potential and promising future research directions to address the gaps. The review comprehensively covers 44 papers recently published in leading transportation journals [1]. Based on the application contexts, the reviewed works are categorized into three groups of strategies: one-lane freeways with a total CAV penetration rate, one-lane freeways with mixed CAV-HDV traffic conditions, and strategies for multilane freeways. We categorize relevant literature based on two criteria: one-lane or multilane in the mainline and CAV-only or mixed traffic flow. The number of lanes of mainline in merging areas determines whether vehicles in mainline can make lane-changing maneuvers and thus significantly influence merging control strategies and problem complexity. CAV-only or mixed traffic determines whether human drivers who cannot be controlled are involved and thus affect control strategy notably. Therefore, control strategies vary pretty remarkably with the two factors, which are the criteria used to categorize relevant literature. The number of reviewed papers for each category is summarized in Table 1.

___

[1] The search covers relevant papers published since 2017 in the following journals: Computer-Aided Civil and Infrastructure Engineering, IEEE Vehicular Technology Magazine, Transport Reviews, Transportation Research Part C: Emerging Technologies, Vehicular Communications, Transportation Research Part E: Logistics and Transportation Review, IEEE Transactions on Intelligent Transportation Systems, IEEE Transactions on Vehicular Technology, Transportation Research Part B: Methodological, Transportation Research Part A: Policy and Practice, Transportation Research Part D: Transport and Environment, Transportation, Journal of Intelligent Transportation Systems, Transportation Science, Transportation Letters, Journal of Intelligent and Connected Vehicles, IEEE Intelligent Transportation Systems Magazine, IEEE Transactions on Intelligent Vehicles, and Journal of Transportation Engineering, Part A: Systems, as well as other highly cited articles in the relevant field.

**Table 1** Number of papers reviewed for different ramp merging categories

| Category No. | Vehicle Technology | Mainline | No. of Reviewed Papers | Primary Methods |
|---|---|---|---|---|
| 1 | CAV-Only | One Lane | 24 | - Optimization<br>- Machine learning<br>- Traffic modeling<br>- Feedback control<br>- Virtual vehicle<br>- Generic algorithm |
| 2 | Mixed | One Lane | 10 | - Optimization<br>- Machine learning<br>- Traffic modeling<br>- Car following<br>- Virtual |
| 3 | CAV-Only or Mixed | Multilane | 10 | - Optimization<br>- Machine learning<br>- Feedback control<br>- Fuzzy control<br>- Game theory |

Note that not all the reviewed studies assume the same level of CAV capabilities. For example, some strategies require Connected vehicles (CV) with on-board driver advisory systems, and some apply to autonomous vehicles (AV) with high-resolution sensors. However, these strategies are essentially the same as the strategies requiring complete CAV capabilities. For example, the strategies that involve only CV are usually achieved by assuming that the human drivers will strictly follow the recommendations from the advisory system (i.e., the human drivers will perform the same level of vehicle control as AV). Therefore, such strategies can be directly transferred to the CAV operation for even better performance. In this sense, we include these studies in the review. Detailed characteristics of the strategies, including vehicle capabilities, penetration level, freeway layout, control decisions, and primary analysis method, are presented in Table 2.

The remainder of the paper is structured as follows. Section 2 reviews the primary strategies for single-lane freeways with a total CAV penetration rate. Section 3 presents the strategies for single-lane freeways with mixed traffic, and Section 4 summarizes the strategies for multilane freeway configurations. Section 5 presents a critical discussion on existing research gaps and future research directions. Finally, Section 6 presents the concluding remarks.

**2 Merging Control for Single-lane Freeways with CAV-Only**

The CAV's ability to perform cooperative and coordinative merging is endowed by emerging vehicle communication and automation technologies. The communication technologies allow for detailed and timely information exchange among road users, traffic infrastructures, and control centers using dedicated short-range radio communications and cellular networks. As a result, the maneuvers of vehicles can be planned via real-time negotiations among traffic participants. Further, autonomous driving systems in vehicles can execute the planned activities can be executed in a stable and timely manner, as they are less prone to delays and errors in the processes of recognition, decision-making, and performance. Depending on the level of control, the existing CAV-enabled merging strategies can be divided into operational control and tactical control. The operational control layer determines lower-level actions of vehicles, such as the step-by-step acceleration and deceleration. In contrast, the tactical control layer addresses the upper-level decisions, such as merging sequence and gap. This section reviews the strategies for the context of the single-lane freeways and total CAV penetration rate (demonstrated in Figure 1) at both levels. In reality, there are generally multiple lanes in the mainline. Therefore, the controls for sing-lane freeways assume no lane-changing behaviors in the most-outer lane.

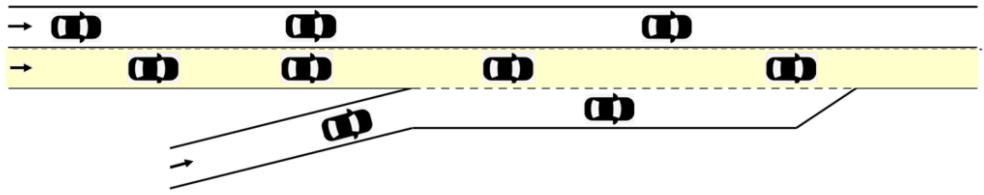

**Figure 1** Merging scenario of single-lane freeways with a total CAV penetration rate.

2.1 Lower-level Control

At the lower level, the actions of relevant vehicles can be planned to enhance the traffic performance at on-ramp merging. A typical approach to solve the mainline-ramp conflict at merging is to utilize the concept of "virtual vehicle/virtual platooning", namely mapping the mainline and ramp vehicles to each other's lane so that the merging problem is transformed into a virtual car-following problem. By following the virtual leader, instead of the physical leader, the vehicles adjust their longitudinal positions in advance for smooth and safe merging. Milanés et al. (2010) develop a fuzzy controller for vehicle throttle and brake control, which allows the vehicles to maintain a reference distance to their virtual leaders. The controller is validated in real-world experiments, where a test vehicle successfully merges in between the other two vehicles. Several subsequent studies use a similar idea of the virtual platoon for collision-free merging operations (Wang et al., 2013, Chen et al., 2021b, Hu et al., 2021, Liao et al., 2021, Mu et al., 2021). Though these approaches facilitate smooth and safe merging, their benefits in merging efficiency are somewhat limited, as the solutions are more intuitive than systematically optimal.

Instead of using intuitive control strategies in the aforenoted literature, other studies were committed to improving CAV merging trajectories under an optimization framework. The key differences of this stream of studies are that they formulated optimization control models and corresponding solvers to obtain the control variables via optimization rather than pre-defined rules. The optimization models in different studies target different objectives favoring traffic efficiency, energy use, and passenger comfort while subject to vehicle dynamics, safety requirements, and technical constraints. For example, Cao et al. (2015) abstracted a ramp merging area as lines in a two-dimensional rectangular coordinate system to describe the interaction between a ramp merging vehicle and its mainline follower. The motions of the vehicles were jointly planned by minimizing a penalty function consisting of acceleration, speed deviation, ramp vehicle's lateral position, and distance between the two conflicting vehicles. The algorithm was implemented in a Model Predictive Control (MPC) scheme and realized cooperative and collision-free merging behaviors in the simulation. Similarly, Zhou et al. (2019a) formulated cooperation between a pair of the ramp and mainline vehicles as two optimal trajectory planning problems related to each other. The algorithm considered explicit bounds on vehicle acceleration and allowed for a flexible choice of merging location. In the optimization framework, the optimal control of facilitated cars in the mainline is given by

$$\min_{u(t)} \int_0^{t_f} \frac{1}{2}[u(t)^2 + \lambda] dt \tag{1}$$

subject to

$$\dot{\mathbf{x}(t)} = \begin{bmatrix} \dot{x}(t) \\ \dot{v}(t) \end{bmatrix} = \begin{bmatrix} v(t) \\ u(t) \end{bmatrix}$$

$$a_{\min} < u(t) < a_{\max}$$
$$x(0) = x_0, \ v(0) = v_0$$
$$x(t_f) = x_e, v(t_f) = v_e \tag{2}$$

where $t_f$ is the period of the merging process; $x(t)$ and $v(t)$ are state variables, representing the distance and speed of the facilitating vehicle; $x_0$ and $v_0$ are the initial states when merging control starts, $x_e$ and $v_e$ are the anticipated and final states after controls; and $\lambda$ is a constant term for penalizing the duration of the merging process. Meanwhile, the optimal control problem for the on-ramp merging vehicle is formulated as

$$\min_{u(t)} \int_0^{t_f} \frac{1}{2}[u(t)^2]dt + \frac{1}{2}\lambda_1[x(t_f) - x_e^m]^2 + \frac{1}{2}\lambda_2[v(t_f) - v_e^m]^2 \quad (3)$$

subject to

$$\dot{\mathbf{x}}(t) = \begin{bmatrix} \dot{x}(t) \\ \dot{v}(t) \end{bmatrix} = \begin{bmatrix} v(t) \\ u(t) \end{bmatrix}$$

$$a_{min} < u(t) < a_{max}$$

$$x(0) = x_0^m, \ v(0) = v_0^m$$

$$x(t_f) = x_e^m, v(t_f) = v_e^m$$

(4)

where $x_e^m$ and $v_e^m$ are desired location and speed of merging vehicles after controls. $\lambda_1$ and $\lambda_2$ are two constant weighting factors to penalize deviation from desired location and speed. The optimization was recursive to address the external disturbances introduced by the leading vehicle, and validated through numerical experiments for improving merging smoothness and computation cost. Similarly, Xu et al. (2021) shed light on the merging of Hybrid Electric Vehicles (HEV). They jointly decided the speed and torque distribution between the engine and motor of an automated HEV at ramp merging for minimal travel time and energy cost. In the branch of centralized merging control, the merging efficiency of relevant vehicles can be jointly improved. Moreover, Liao et al. (2021) designed a cooperative system for connected vehicles based on Vehicle-to-Cloud (V2C) communication. Vehicles uploaded real-time status to the cloud server to determine the advisory speed for a ramp vehicle to merge between two mainline vehicles.

However, merely considering one vehicle in trajectory planning will neglect the impacts of controls of one vehicle on other vehicles in the traffic flow, which may lead to sub-optimal. To overall such a shortcoming, the control scope was extended from a single pair of vehicles to a series of vehicles in the communication range. Ntousakis et al. (2016) assumed the existence of a pre-determined merging sequence of relevant vehicles, and the vehicles collected information from their physical and putative leaders to plan their optimal trajectories. The study considered various cost functions, including vehicle acceleration, jerk, and the first derivative of the jerk. The results showed its superiority to a typical Adaptive Cruise Control (ACC) controller in terms of engine effort and passenger comfort. Rios-Torres et al. (2017a) organized the vehicles in the first-in-first-out order and established an optimal framework to minimize the total fuel consumption of all vehicles in the control area. Merging collisions were avoided by regulating only one vehicle that may cross the merging zone. A closed-form solution was analytically derived using Hamiltonian analysis, and simulations were carried out to examine the benefits of the merging algorithm in smoothing vehicle trajectory and reducing fuel consumption. Later, the idea to reduce fuel consumption via trajectory planning was further adopted by Sonbolestan et al. (2021). The authors determined the optimal merging position of a ramp vehicle for the mainline. The strategy was tested in a simulation with field traffic data as inputs. Letter et al. (2017) developed an optimization-based strategy to maximize the average speed of vehicles in the merging area. The algorithm first determined a merging sequence based on the estimated arrival times of vehicles and then constrained motions of each vehicle on the trajectory of its leading vehicle. The strategy worked even in oversaturated conditions to increase merging throughput, efficiency, and comfort. Similarly, the strategy in Xie et al. (2017) targeted maximizing the total speed of all vehicles over a specific decision interval while ensuring safety distances between successive vehicles in all lanes. The developed strategy outperformed a no-control case and a gradual speed limit strategy in terms of throughput, vehicle speed, and delay.

Besides the abovementioned methods, some other approaches have been developed for motion decisions at CAV ramp merging. In Ward et al. (2017), a set of candidate trajectories were evaluated to select the optimal one based on a cost function incorporating merging progress, comfort, and risk. The evaluation of candidate trajectories considers the responsive motions of surrounding vehicles predicted by a probabilistic Intelligent Driver Model (IDM). Fukuyama (2020) used a two-player dynamic game approach to interpret the interactions between a competing pair of vehicles. Under this approach, each vehicle made trajectory decisions to maximize its driving utility while considering the potential actions/responses of the competing vehicle. Recently, feedback and feedforward methods have been applied in real-time CAV motion control. For example, Chen et al. (2021b) designed a motion controller that used the speed difference and spacing error as the feedback components and the acceleration of predecessors as the feedforward information to produce longitudinal motion commands. The string stability for the proposed controller was analytically proven. In Hu et al. (2021), a motion controller was designed to handle the non-linearity and time-varying uncertainties in vehicle dynamics. The controller took the spacing error as the control object. Its effectiveness was assessed under critical situations with speed variations of the surrounding vehicles and simultaneous merging of multiple ramp vehicles.

2.2 Upper-level Control

As reviewed in Section 2.1, many lower-level motion planning strategies require a merging sequence as input. However, the merging sequence is determined intuitively based on relatively simple rules, such as first-in-first-out, virtual vehicle mapping, and estimated arrival time. This leaves a chance to further improve the merging coordination through more efficient designs of the merging sequence.

Some recent studies shed light on the problem of merging sequence planning. For example, Xu et al. (2019b) employed a genetic algorithm to solve the choice of a merging sequence. In addition, the utilities of candidate sequences were assessed by a fitness function combining the travel time of mainline vehicles and the number of ramp vehicles allowed to merge, as shown by

$$\max F = w_1 f_{mainline} + w_2 f_{merging} + g_1 e_v + g_2 e_s \tag{5}$$

subject to

$$\begin{aligned} & 0 \leq w_1 \leq 1 \\ & 0 \leq w_2 \leq 1 \\ & w_1 + w_2 = 1 \\ & a_{\min} \leq a_j \leq a_{\max} \end{aligned} \qquad (6)$$

where $f_{mainline}$ and $f_{merging}$ are the utility functions for vehicles in the mainline and ramps, which are related to the travel time and number of merged vehicles; $w_1$ and $w_2$ are constant weights; $e_v$ is the error of vehicle speeds after merging; $e_v$ is the error of spacings between vehicles after merging; and $g_1$ and $g_2$ are penalty factors. A genetic algorithm is used to solve the optimization. The algorithm encoded the merging sequences in a binary representation and samples candidate sequences through selection, crossover, and mutation, as illustrated in Figure 2. After a certain number of generations, the merging sequence with the maximal fitness value was chosen. In Xu et al. (2019a), vehicles whose inter-vehicle distances were lower than a pre-defined threshold were grouped, and the merging order of vehicles within a group remained unchanged. Then, the groups were sequenced by enumerating all possible orders to select the one with minimal total passing time and delay. The proposed strategy could find near-optimal solutions with less computation time and a good tradeoff between computation efficiency and traffic performance. Similarly, Pei et al. (2019) developed a strategy to improve the computation efficiency of searching the optimal merging sequence at the prerequisite of maintaining the optimality of the result. By stipulating that vehicles on the same road follow a first-in-first-out order, the strategy assigned the right of way between the two links (mainline/ramp) instead of among individual vehicles so that the complexity of the sequencing problem was substantially reduced.

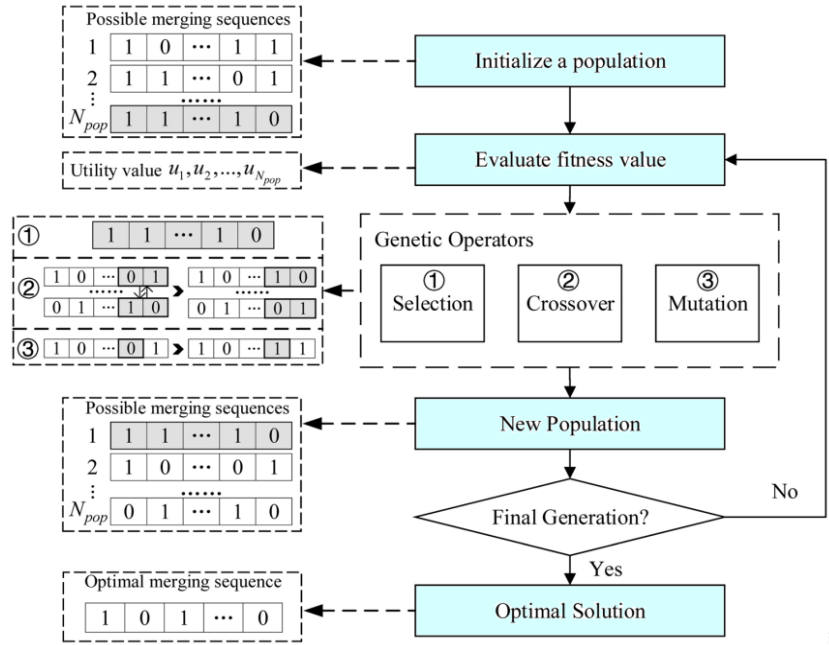

**Figure 2** Merging sequence representation and flowchart of used genetic algorithm (Xu et al. 2019b).

Moreover, several recent studies have integrated the choice of a merging sequence with the planning of vehicle trajectories. For example, Ding et al. (2020) designed a series of rules to adjust the merging order of vehicles and then planned the motion of each vehicle accordingly. Based on the idea that the headway between vehicles from different roads is usually larger than that from the same road, the proposed rules allowed vehicles from the same road to pass the merging area together to reduce the switch of the right of way. Jing et al. (2019) developed an optimization model for integrated merging sequence and trajectory decisions. The model objective consisted of a strategy cost related to merging sequence and an action cost determined by vehicle accelerations and jerks. Alternative to a decision on the merging sequence, Chen et al. (2020) determined a merging gap for each ramp vehicle. The strategy was composed of a tactical layer and an operational layer. The tactical layer used a second-order dynamics model to estimate the trajectory costs for different gaps and select the gap with the least cost for the vehicle. The operational layer further optimized the trajectory with a third-order dynamics model. Similarly, Nishi et al. (2019) made a joint decision on the merging gap and its trajectory for each ramp vehicle. The strategy chose the policy with the lowest cost based on the state value function learned from field data using a passive actor-critic method.

The above CAV ramp merging strategies mainly focused on controlling individual vehicles and benefits at the local and microscopic levels. Through an aggregated control of the two streams of traffic at on-ramp merging, improvements in the overall traffic flow performance can be achieved. An example of such an aggregated control system was introduced in Scarinci et al. (2015) and Scarinci et al. (2017). Under the control strategy, the mainline traffic was periodically compacted to create large gaps, and the ramp vehicles were released into the gaps through ramp metering signals. However, this strategy stipulated that the release of ramp vehicles entirely depended on the mainline conditions, so the efficiency of ramp traffic was not actively considered. The strategy was established based on the macroscopic traffic flow theories, with explicit consideration of the dynamics of gap formation. The validation indicated that the proposed strategy was compatible with real-world implementation and could reduce the occurrence of congestions and the number of late-merging vehicles.

## 3 Merging Control for Single-lane Freeways with Mixed Traffic

It is widely predicted that before the complete launch of CAVs, there will be a long period when the CAVs and HDVs share the public roads. This will lead to significant challenges in the ramp merging control of CAVs, as the uncontrollability of HDVs will introduce various uncertainties into traffic operation. Therefore, recent research efforts are applied to the design of CAV control strategies in the presence of HDVs, as displayed in Figure 3. This section will review control strategies in the contexts of single-lane freeways with both CAVs and HDVs.

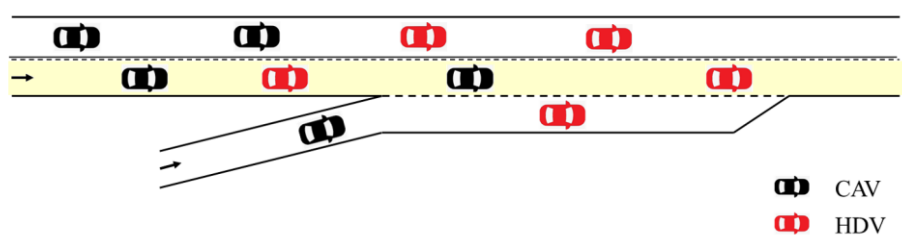

**Figure 3** Merging scenario of single-lane freeways with mixed traffic.

A typical control problem for the mixed traffic is to guide a CAV from the on-ramp to merge into the mainline traffic with HDVs, where the actions of HDVs should be predicted and explicitly considered in the motion plan of CAVs. To resolve this issue, Kherroubi et al. (2021) trained a probabilistic classifier using artificial neural networks and field data to predict the passing intentions of human drivers. The prediction further served as an input of a reinforcement learning agent to control the longitudinal acceleration of a ramp CAV. The simulation showed the control's ability to reduce the number of vehicle conflicts and stops at ramp merging. In Okuda et al. (2021), a logistic regression model was developed to estimate the decisions of the mainline human drivers (expressed as the probability to accept a vehicle to merge in front of it). Based on these decisions, a merging strategy was proposed to maximize the acceptance probability of the mainline human drivers by proactively adjusting the speed of the merging CAV. The optimization formulation is given by.

$$\min \sum_{k=1}^{K} \sum_{c=1}^{N} S^c(k \mid t) \quad (7)$$

subject to

$$S^c(t) = -\sum_{s=1}^{3} P(X_{SOD}(t) = s \mid \phi(t)) \times \log_2 P(X_{SOD}(t) = s \mid \phi(t))$$

$$P(X_{SOD}(t) = s \mid \phi(t)) = \begin{cases} \dfrac{\exp(\eta_s^T \phi(t))}{1+\exp(\eta_1^T \phi(t))+\exp(\eta_2^T \phi(t))} & \text{if } s = 1 \text{ or } 2 \\ 1 - P_1(\phi(t)) - P_2(\phi(t)), & \text{if } s = 3 \end{cases}$$

$$\mathbf{A}^E(t) = (d^{M,E}(t), v^{M,E}(t), a^{M,E}(t))^T \quad (8)$$

$$\mathbf{E}^E(t) = (d^{L,E}(t), v^{\gamma,E}(t), L_w)^T$$

$$\phi(t) = (1, \mathbf{A}^E(t)^T, \mathbf{E}^E(t)^T)$$

$$s = 1 : accept$$

$$s = 2 : rejected$$

$$s = 3 : undecided$$

where $t$ is the current time and $k$ is the time index in the prediction horizon; $K$ is the total number of steps in the prediction horizon; $d^{M,E}$, $v^{M,E}$ and $a^{M,E}$ are distance, relative speed and relative acceleration between the following vehicle in the mainline and merging vehicle in ramps, respectively; $d^{L,E}(t)$ is the distance between the leading vehicle in the mainline and the merging vehicle and $v^{\gamma,E}(t)$ is the distance of the merging vehicle to the end of the acceleration lane. The control strategy was executed in a model predictive control scheme. In contrast to the control of ramp merging CAVs, Zhou et al. (2017) focused on the control of CAVs on the main road. The authors used the intelligent driver model to capture the adaptive and cooperative behaviors of the mainline CAVs. Further, they showed through an illustrative case study that the increase in CAV penetration rate could reduce the total travel time and smooth traffic oscillations at on-ramp merging.

Moreover, some research was devoted to facilitating ramp merging through simultaneous control of the mainline and ramp vehicles. The earlier work by Zhou et al. (2019a) jointly planned motions of a ramp merging vehicle and a mainline facilitating vehicle. Subsequently, Zhou et al. (2019b) further introduced a lower bound on the cooperative speed of the facilitating vehicle to restrain the adverse impacts on the upstream traffic and avoid undesired speed-drops on the main road. The improved strategy was tested under mixed flow conditions. The results showed the strategy's ability to reduce the risk of rear-end collisions and mitigate speed variations of the mainline traffic flow. Karimi et al. (2020) divided the merging situations into six categories depending on the combinations of CAVs and HDVs in a merging triplet (i.e., a merging vehicle and its putative leader and follower in the target lane). They also developed for each category a cooperative strategy that checked the desired speed and inter-vehicle distance at a series of set-points. Numerical studies showed that such a strategy could achieve smooth and cooperative merging.

The above strategies for mixed traffic flow focused on the interaction between a single ramp vehicle and its direct neighbors on the main road. However, they neglected the influence on the surrounding traffic. By jointly planning the motions of multiple vehicles under an optimization framework, the overall traffic performance in the merging area can be improved. Mu et al. (2021) developed a trajectory planning method that first forms multiple mainline and ramp vehicles into a virtual platoon. Then, they planned their motions to maximize the average vehicle distance traveled over a specific planning period. The strategy was extended to the mixed CAV-HDV conditions in two steps: (1) dividing the mixed string of vehicles into blocks containing a leading CAV and several following HDVs, (2) considering each block as a whole trajectory planning problem. Ito et al. (2019) designed a merging planning system where a global controller collects CAV information and estimates the states of HDVs. The state information was encoded and broadcast to all road users in the merging areas so that the local controllers on CAVs could use the information to plan their trajectories. Omidvar et al. (2020) extended the merging strategy in Letter et al. (2017) for the mixed traffic flow. The model accounted for deviations between the predicted and actual behaviors of HDV through a real-time correction mechanism yet oversimplified HDV driving patterns. Sun et al. (2020) integrated the choice of merging gap and the design of vehicle trajectories into an optimization problem. The problem assumed different driving rules for CAVs and HDVs and considered the benefits of a series of upstream vehicles on the main road. Ding et al. (2019) applied their previous strategy (Ding et al., 2020) to the mixed CAV-HDV condition. They discussed the impacts of CAV penetration on throughput, traffic efficiency, fuel use, and emission. Similarly, Rios-Torres et al. (2018) applied the motion planner developed in Rios-Torres et al. (2017a) to an environment where CAV and HDV coexisted to investigate how the increasing percentage of CAVs may influence the energy use at on-ramp merging. Nevertheless, these studies (Rios-Torres et al., 2018, Ding et al., 2019) were more focused on assessing the impacts of CAV penetration under the existing control framework rather than designing a specific CAV control strategy for the mixed traffic conditions.

At the traffic flow level, Chen et al. (2021a) adopted the idea of periodic gap creation and combined it with a batch merging strategy to close the extra time gaps induced by the lane-changing maneuvers of on-ramp vehicles. It was demonstrated in theory that the proposed system could reduce unutilized roadway capacity and increase merging throughput, but no numerical/simulation experiment carried out.

**4 Merging Control for Multilane Freeways**

The above-reviewed strategies consider a simple freeway layout with one lane on each road and ignore the lane-changing behaviors between the mainline lanes. Nevertheless, the single-lane freeway layout is less common in practice. This section reviews the CAV ramp merging strategies designed for the more realistic multilane freeway configurations, as demonstrated in Figure 4. With the presence of multiple lanes on the main road, the lane-changing decisions of mainline vehicles may influence the merging opportunities of ramp vehicles and introduce more uncertainties in the ramp merging cooperation, presenting significant challenges for CAV control.

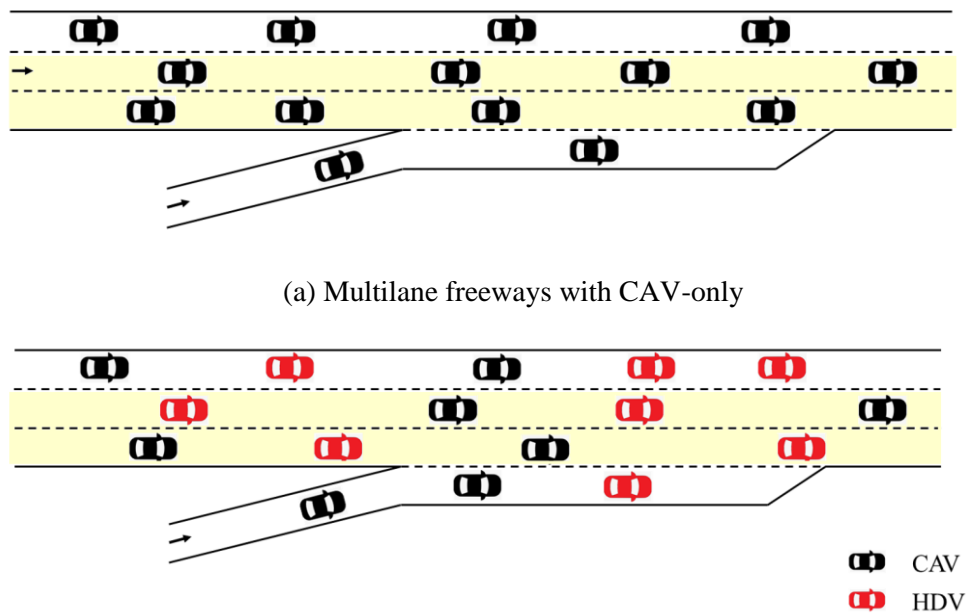

(a) Multilane freeways with CAV-only

(b) Multilane freeways with mixed traffic

**Figure 4** Merging scenario of multilane freeways.

A prior effort to address such a challenge is Marinescu et al. (2012). The authors designed a hybrid system combining centralized and decentralized control to account for the negotiations between the mainline and ramp vehicles. The system divided the space on a multilane freeway into moving slots and assigned slots to vehicles for easy merging. It was stipulated that the mainline vehicles tend to move into the free slots in their front left to utilize the space in the inner lanes so that more slots were released for the merging of on-ramp vehicles. Following similar ideas, subsequent studies proposed various solutions to release the space in the outermost lane via proactive controls of CAVs. For example, Karbalaieali et al. (2020) considered a two-lane freeway and chose from various combinations of alternative actions (i.e., speeding up, slowing down, or changing lanes) the one that minimized the total travel time of a ramp vehicle and its direct mainline competitors. Though presenting improvements in traffic efficiency, the proposed strategy employed a discrete decision space, leaving room for more sophisticated action control. Hang et al. (2021) interpreted the merging process as a coalitional game involving a ramp vehicle and the mainline vehicles directly influenced by it. In the game, each vehicle decided whether it would behave cooperatively (join a coalition) or independently (leave a coalition) based on its individualized orientation towards efficiency, safety, and comfort. The lane-changing decisions and longitudinal vehicle trajectories in each coalition were optimized for the maximal benefits of the coalition. Karbalaieali et al. (2020) and Hang et al. (2021) considered one ramp vehicle only once, making the strategies less effective when the on-ramp vehicles arrived frequently. To accommodate the merging of multiple ramp vehicles, Hu et al. (2019) extended the trajectory optimization strategy of Letter et al. (2017) to implement a multilane layout. Their extensive strategy designed a cooperative lane-changing zone upstream of the trajectory control zone. A part of mainline vehicles in the outer lane was allocated to the inner lane to balance the after-merging flow between mainline lanes. The merging trajectory was optimized to maximize the subject vehicle's velocity during the optimization process, as follows

$$\min(-\sum_{t=1}^{T_{CM}} v_t) \qquad (9)$$

subject to

$$\begin{aligned}
&v_t \in [0, v_{\max}^{CM}] \\
&a_t \in [a_{\min}^{CM}, p \cdot v_t + q] \\
&|j_t| \leq j_{\max}^{CM} \quad \forall t \in \{2,3,...,T^{CM}\} \\
&X_t - x_t \geq v_t g_{\min}^{CM} \\
&X_t - x_t \geq d_{\min} \\
&x_t - x_{t-1} = v_{t-1} \Delta t^{CM} \\
&v_t - v_{t-1} = a_{t-1} \Delta t^{CM}
\end{aligned} \qquad (10)$$

where $t$ is time step index; $x_t$ and $X_t$ are the positions of the subject vehicle and conflicting vehicle, respectively; $v_t$, $a_t$ and $j_t$ are velocity, acceleration, and jerk of the subject vehicle; $g_{\min}^{CM}$ and $d_{\min}$ are the minimum time headway and distance headway during merging; $\Delta t^{CM}$ is the merging length of the time interval. The simulation showed that the comprehensive strategy outperformed the original strategy to reduce delay and increase vehicle speed. Liu et al. (2021) also integrated a lane selection model with the trajectory planning problem to account for the unevenness between lanes. The lane selection model employed a reinforcement learning approach to output the lane choice decision of each vehicle based on the real-time traffic flow conditions.

Moreover, recent studies shed light on the more complicated situations where CAVs and HDVs coexist in a multilane merging area. Gao et al. (2021) developed an optimization-based trajectory planning strategy that enabled flexible choices of the mainline facilitating vehicle and the merging point. The strategy was tested under various CAV penetration rates in a two-lane freeway merging area. Williams et al. (2021) considered the challenging situation where vehicles may freely change lanes on a multilane freeway, especially with the presence of uncontrolled HDVs, so that the coordinated mainline vehicles may vary over time. The study proposed a mechanism to update the merging sequence and accounted for deviations in the measured vehicle positions in real-time to address this issue. Guo et al. (2020) developed a reinforcement learning approach for lane change control at freeway on-ramps and off-ramps. The automated agents took the speed of a vehicle and its distances to the surrounding vehicles (in the current and adjacent lanes) as inputs and output the decisions of speed (increase/decrease speed) and lane-change (keep the current lane or change). Recent studies considered the macroscopic traffic flow performance and proposed strategies that integrate multiple control measures at on-ramps. For example, Tajdari et al. (2020) combined lane-changing control enabled by CAVs with the conventional ramp metering strategy. Based on a cell-based traffic flow model, the combined strategy sought to keep the bottleneck density at the critical density level for maximized throughput by deciding the lane-changing flows between lanes and the ramp inflow rate. Adopting a similar idea, Pan et al. (2021) integrated ramp metering, variable speed control, and lane change control for CAVs and the corresponding recommendations to HDVs in a merging control system. The strategy explicitly considers the stability of traffic flow and the compliance rate of human drivers. Numerical experiments show clear benefits of the strategy in terms of traffic efficiency, energy use, and emissions.

## 5 Future research directions

Based on the comprehensive review of existing studies in the field of CAV ramp merging coordination, the following research gaps are identified:

- Most previous studies of CAV ramp merging have focused on the lower-level design of vehicle trajectories. In contrast, some have considered the upper-level decisions, such as selecting merging sequence and merging gap. In addition, only a few studies have addressed the development of control strategies at the traffic flow level. Therefore, a comprehensive system considering various essential elements of both control levels is needed.

- The existing strategies for mixed traffic conditions usually make simple assumptions on the driving patterns of HDVs. For example, the strategies assume that the HDVs strictly follow specific predefined driving rules without errors, delays, or variances in the human driver population. Unfortunately, these assumptions tend to underestimate the uncertainties induced by HDVs and overestimate the cooperation willingness of human drivers.

- Most existing strategies for mixed traffic flow regard HDVs as only an uncontrolled external factor that restricts CAV behaviors, while the possibilities to influence the behaviors of HDVs for enhanced coordination benefits are not explored.

- As the multilane control problem is relatively complicated, many existing strategies only propose solutions in a discrete decision space (e.g., increase/decrease one step in speed and change lane or not). Therefore, more sophisticated coordination strategies in both the longitudinal and lateral directions, based on a thorough analysis of the consequences at the traffic flow level, should be further investigated.

## 6. Concluding Remarks

This paper has presented a comprehensive review of the existing freeway ramp merging strategies leveraging CAVs, focusing on the latest trends and developments in the field. Based on the application context, the strategies are divided into three categories: (1) strategies involving CAV-only merging into one lane, (2) strategies for mixed traffic flow merging into one lane, and (3) strategies for CAV-only or mixed traffic flow merging into multi lanes. The findings based on the review can be summarized as follows:

Prior studies on CAV merging coordination tend to simplify the merging context and assume single-lane freeway layouts with a total CAV penetration rate. In the simplified contexts, the free lane-changing behaviors on the main road and the uncertainties induced by the uncontrolled HDVs are ignored. These considerations are continually accommodated in recent research efforts. Studies considering the presence of HDVs are devoted to predicting the intentions/behaviors of HDVs and use the predictions as inputs for the CAV control plan. In multilane freeway configurations, proactive lane-changing decisions of mainline CAVs are included in the control framework to fully utilize the bottleneck capacity and increase the overall performance of ramp merging operation.

Optimization is the most widely used method to solve the CAV merging problem, especially for the trajectory planning problem. The objectives of the optimization models favor different performance indexes, such as traffic efficiency, safety, energy use, and passenger comfort, while being subject to constraints on vehicle dynamics, technical limits, safety constraints, and requirements on traffic flow operation. Recent studies explored the possibilities of applying other methods to CAV merging control, such as machine learning, game theory, and feedback control. A combination of multiple methods is also used in a few studies (Guo et al., 2020, Hang et al., 2021, Liu et al., 2021).

Most of the existing studies of CAV ramp merging emphasize lower-level controls such as trajectory planning. Even though some studies addressed upper-level controls for one-lane freeways with CAV-only, very limited studies have conducted optimization for upper-level and traffic flow controls for multilane merging and one-lane freeways with mixed traffic, which are the reasons why we do not review upper-level and lower-level controls separately for one-lane freeways with mixed traffic in Section 3 and multilane merging in Section 4. Therefore, more efforts in these streams should be made to improve robust on-ramp merging controls for multilane configurations and mixed traffic that are rather common in the forthcoming future.

Despite the comprehensive review presented in this paper, the study has some limitations. First, the review focuses on recent developments in CAV ramp merging coordination. However, some earlier studies regarding automated highway systems (Ioannou 2013, Easa 2007), for example, are not included in the scope of the review. Instead, the reader is referred to Scarinci et al. (2014) and Rios-Torres et al. (2017b) for an excellent review of prior works. Second, this review focuses on the CAV on-ramp merging situation studies. However, further studies covering the strategies for comparable situations, such as merging at work zones (Cao et al., 2021), lane drops (Zhang et al., 2019), and lane change at off-ramps and weaving sections (Zheng, 2014, Amini et al., 2021, Nagalur Subraveti et al., 2021) may benefit the overall understanding of CAV merging problem in general. Therefore, a survey including such studies is part of our ongoing review plan.

**Acknowledgement**

The authors are grateful to VINNOVA (ICV-Safe, 2019-03418), Area of Advance Transport, and AI Center (CHAIR) at the Chalmers University of Technology for funding this research.

**Table 2** Characteristics of the reviewed ramp merging strategies

| Category | Reference | Veh. Tech. | Penet. Level | Freeway Layout | Control Decision | | Primary Method |
|---|---|---|---|---|---|---|---|
| | | | | | **Upper Level** | **Lower level** | |
| 1. CAV-only and One-lane | Chen et al. (2021b) | CAV | Total | One-lane | - [b] | Trajectory | Feedback control |
| | Hu et al. (2021) | CAV | Total | One-lane | - | Trajectory | Feedback control |
| | Liao et al. (2021) | CV | Total | One-lane | - | Trajectory | Feedback control |
| | Sonbolestan et al. (2021) | CAV | Total | One-lane | - | Trajectory | Optimization |
| | Xu et al. (2021) | CAV | Total | One-lane | - | Trajectory | Optimization |
| | Chen et al. (2020) | CAV | Total | One-lane | Merging gap choice | Trajectory | Optimization |
| | Ding et al. (2020) | CAV | Total | One-lane | Merging sequence | Trajectory | Optimization |
| | Fukuyama (2020) | CAV | Total | One-lane | - | Trajectory | game theory |
| | Jing et al. (2019) | CAV | Total | One-lane | Merging sequence | Trajectory | Optimization |
| | Nishi et al. (2019) | AV | Total | One-lane | Merging gap choice | Trajectory | Machine learning |
| | Pei et al. (2019) | CAV | Total | One-lane | Merging sequence | - | Optimization |
| | Xu et al. (2019a) | CAV | Total | One-lane | Merging sequence | - | Optimization |
| | Xu et al. (2019b) | CAV | Total | One-lane | Merging sequence | Trajectory | Generic algorithm |
| | Zhou et al. (2019a) | CAV | Total | One-lane | - | Trajectory | Optimization |
| | Zhou et al. (2019b) | CAV | Total | One-lane | - | Trajectory | Optimization |
| | Letter et al. (2017) | CAV | Total | One-lane | - | Trajectory | Optimization |
| | Rios-Torres et al. (2017a) | CAV | Total | One-lane | - | Trajectory | Optimization |
| | Scarinci et al. (2017) | CAV | Total | One-lane | Periodic gap | - | Traffic modeling |
| | Ward et al. (2017) | AV | Total | One-lane | - | Trajectory | Optimization |
| | Xie et al. (2017) | CAV | Total | One-lane | - | Trajectory | Optimization |
| | Ntousakis et al. (2016) | CAV | Total | One-lane | - | Trajectory | Optimization |
| | Cao et al. (2015) | CV | Total | One-lane | - | Trajectory | Optimization |
| | Wang et al. (2013) | CAV | Total | One-lane | - | Trajectory | Virtual vehicle |
| 2. Mixed and One-lane | Chen et al. (2021a) | CAV | Mixed | One-lane | Periodic gap, batch merge | - | Traffic modeling |
| | Rios-Torres et al. (2018) | CAV | Mixed | One-lane | - | Trajectory | |
| | Kherroubi et al. (2021) | CAV | Mixed | One-lane | - | Trajectory | Machine learning |
| | Mu et al. (2021) | CAV | Mixed | One-lane | - | Trajectory | Optimization |
| | Okuda et al. (2021) | AV | Mixed | One-lane | - | Trajectory | Optimization |
|  | Karimi et al. (2020) | CAV | Mixed | One-lane | - | Trajectory | Optimization |
|  | Omidvar et al. (2020) | CAV | Mixed | One-lane | - | Trajectory | Optimization |
|  | Sun et al. (2020) | CAV | Mixed | One-lane | Merging gap choice | Trajectory | Optimization |
|  | Ding et al. (2019) | CAV | Mixed | One-lane | Merging sequence | Trajectory | Optimization |
|  | Ito et al. (2019) | CAV | Mixed | One-lane | - | Trajectory | Optimization |
|  | Zhou et al. (2017) | AV | Mixed | One-lane | - | Trajectory | Car-following |
| 3. CAV-only or Mixed Multilane | Gao et al. (2021) | CAV | Mixed | Multilane | - | Trajectory | Optimization |
|  | Hang et al. (2021) | CAV | Total | Multilane | - | LC, trajectory | Game theory |
|  | Pan et al. (2021) | CAV | Mixed | Multilane | Ramp inflow | LC, speed | Optimization |
|  | Liu et al. (2021) | CAV | Total | Multilane | - | LC, trajectory | Machine learning |
|  | Williams et al. (2021) | CAV | Mixed | Multilane | - | Trajectory | Car-following |
|  | Guo et al. (2020) | CAV | Mixed | Multilane | - | LC[a], speed | Machine learning |
|  | Karbalaieali et al. (2020) | CAV | Total | Multilane | - | LC, speed | Optimization |
|  | Tajdari et al. (2020) | CV | Mixed | Multilane | LC/ramp flows | - | Feedback control |
|  | Hu et al. (2019) | CAV | Total | Multilane | - | Trajectory | Optimization |
|  | Milanés et al. (2010) | CV | Total | Multilane | - | Trajectory | Fuzzy control |

[a] LC = Lane change. [b] Not applicable.